\documentclass[pre,twocolumn,twoside,byrevtex,superscriptaddress,floatfix]{revtex4}

\usepackage{url}
\usepackage{graphicx,epsfig,verbatim,enumerate}
\usepackage{amssymb,amsmath}
\usepackage{ifthen}
\newboolean{twocolswitch}

\usepackage{color}

\definecolor{lightgrey}{rgb}{0.5,0.5,0.5}

\usepackage{rotating}
\usepackage{longtable}
\usepackage{array}

\newcommand{\sindex}[1]{}
\newcommand{\nindex}[1]{}

\newcommand{\www}[1]{\url{#1}}

\newcommand{\PreserveBackslash}[1]{\let\temp=\\#1\let\\=\temp}
\newcommand{\PBS}[1]{\let\temp=\\#1\let\\=\temp}

\newcommand{\havgword}[1]{h_{\rm avg}(\mbox{`#1'})}

\newcommand{\havgfn}{h_{\rm avg}}

\setboolean{twocolswitch}{true}

\newcommand{\revtexonly}[1]{#1}
\newcommand{\plainlatexonly}[1]{}

\begin{document}

\title{
  Positivity of the English language\revtexonly{\\ \small PLoS ONE, Vol 7, e29484, 2012}

}

\author{
\firstname{Isabel M.}
\surname{Kloumann}
}
\email{isabel.kloumann@uvm.edu}

\author{
\firstname{Christopher M.}
\surname{Danforth}
}
\email{chris.danforth@uvm.edu}

\author{
\firstname{Kameron Decker}
\surname{Harris}
}
\email{kameron.harris@uvm.edu}

\author{
\firstname{Catherine A.}
\surname{Bliss}
}
\email{catherine.bliss@uvm.edu}

\author{
\firstname{Peter Sheridan}
\surname{Dodds}
}
\email{peter.dodds@uvm.edu}

\affiliation{
  Department of Mathematics and Statistics,
  Center for Complex Systems,
  \&
  the Vermont Advanced Computing Center,
  University of Vermont,
  Burlington,
  VT, 05401
}

\date{\today}

\begin{abstract}
   Over the last million years,
human language has emerged and evolved
as a fundamental instrument of
social communication and semiotic representation.
People use language in part to convey emotional information,
leading to the central and contingent questions:
(1) What is the emotional spectrum of natural language?
and 
(2) Are natural languages neutrally, positively, or negatively biased?
Here, we report that the human-perceived positivity
of over 10,000 of the most frequently used English words 
exhibits a clear positive bias.
More deeply, we characterize and quantify distributions of word positivity
for four large and distinct corpora, 
demonstrating that
their form is broadly invariant with respect to frequency of word use.

\end{abstract}

\maketitle

\section*{Introduction}

While we regard ourselves as social animals,
we have a history of actions running from selfless
benevolence to extreme violence at all scales of society,
and we remain scientifically and philosophically unsure
as to what degree any individual or group is or should be 
cooperative and pro-social.
Traditional economic theory of human behavior,
for example, assumes that people
are inherently and 
rationally selfish---a core attribute of \textit{homo economicus}---with the emergence
of global cooperation thus rendered a profound mystery~\cite{axelrod1984a,nowak2006a}.
Yet everyday experience and many findings
of psychology, behavioral economics,
and neuroscience indicate people
favour seemingly irrational heuristics~\cite{richerson2005a,ariely2010a}
over strict rationality
as exemplified in loss-aversion~\cite{kahneman1990a}, 
confirmation bias~\cite{nickerson1998a},
and altruistic punishment~\cite{fehr2002a}.
Religions and philosophies similarly run the gamut in
prescribing the right way for individuals to behave,
from the universal non-harming advocated by Jainism,
Gandhi's call for non-violent collective resistance,
and exhortations toward altruistic behavior in all major religions,
to arguments for the necessity of a Monarch~\cite{hobbes1651a},
the strongest forms of libertarianism, 
and the ``rational self-interest'' of Ayn Rand's Objectivism~\cite{rand1964a}.

In taking the view that humans are in part story-tellers---\textit{homo narrativus}---we 
can look to language itself for quantifiable 
evidence of our social nature.  
How is the structure of the emotional content rendered in our stories, fact or fiction, 
and social interactions reflected in the collective, evolutionary 
construction of human language?
Previous findings are mixed: suggestive evidence of a positive bias has been found in
small samples of English words~\cite{boucher1969a,rozin2010a,augustine2011a},
framed as the Pollyanna Hypothesis~\cite{boucher1969a} and Linguistic Positivity Bias~\cite{augustine2011a},
while experimental elicitation of emotional words 
has instead found a strong negative bias~\cite{schrauf2004a}.

To test the overall positivity of the English language,
and in contrast to previous work~\cite{bradley1999a,schrauf2004a,rozin2010a},
we chose words based solely on frequency of use,
the simplest and most impartial gauge of word importance.
We focused on measuring happiness, or psychological valence~\cite{osgood1957a}, 
as it represents the dominant emotional response~\cite{chmiel2011a,reisenzein1992a}.
With this approach, we examined four large-scale text corpora
(see Tab.~\ref{tab:wordhap.corpora} for details):
Twitter,
The Google Books Project (English),
The New York Times, and
Music lyrics.
These corpora, which we will refer to as
TW, GB, NYT, and ML,
cover a wide range of written expression
including broadcast media, opinion, 
literature, songs,
and public social interactions (\cite{kwak2010a}),
and span the gamut in terms of grammatical
and orthographic correctness.

\begin{table*}[thp!]
  \centering
  \begin{tabular}{|l|l|c|c|c|}
    \hline
    \hline
    Corpus (Abbreviation): 
    & 
    Date range 
    & 
    \# Words 
    & 
    \# Texts 
    &
    Reference
    \\
    \hline
    Twitter (TW)
    &
    9/9/2008 to 3/3/2010
    &
    9.07$\times$$10^9$ 
    &
    8.21$\times$$10^8$ tweets
    & 
    \cite{twitterapi,dodds2011e}
    \\
    Google Books Project, English (GB)
    &
    1520 to 2008
    &
    3.61$\times$$10^{11}$
    &
    3.29$\times$$10^6$ books
    & 
    \cite{googlebooks-ngrams,michel2011a}
    \\
    The New York Times (NYT)
    &
    1/1/1987 to 6/30/2007
    &
    1.02$\times$$10^9$
    & 
    1.8$\times$$10^6$ articles
    & 
    \cite{nytimescorpus2008a} 
    \\
    Music lyrics (ML)
    &
    1960 to 2007
    & 
    5.86$\times$$10^7$
    & 
    2.95$\times$$10^5$ songs
    & 
    \cite{dodds2009b}
    \\
    \hline
    \hline
  \end{tabular}
  \caption{
    Details of the four corpora we examined for positivity bias.
  }
  \label{tab:wordhap.corpora}
\end{table*}

We took the top 5000 most frequently used words from
each corpus, and merged them to form a resultant list of 10,222 unique words.
We then used Amazon's Mechanical Turk~\cite{mechturk,dodds2011e} to obtain 50 independent evaluations per word
on a 1 to 9 integer scale, asking participants to rate their happiness 
in response to each word in isolation 
(1 = least happy, 5 = neutral, and 9 = most happy~\cite{bradley1999a,dodds2009b}).
While still evolving, Mechanical Turk has proved 
over the last few years to be a reliable and fast service
for carrying out large-scale social science
research~\cite{snow2008a,miller2011a,bohannon2011c,paolacci2010a,rand2011a}.

We computed the average happiness score and standard deviation for each word.
We obtained sensible results that showed excellent
statistical agreement with previous studies for smaller word sets,
including a translated Spanish version 
(see \cite{bradley1999a,dodds2011e,redondo2007a} for details).
The highest and lowest scores were 
$\havgword{laughter}$=8.50
and
$\havgword{terrorist}$=1.30,
with expectedly neutral words averaging near 5,
e.g., $\havgword{the}$=4.98
and
$\havgword{it}$=5.02.
We refer to our ongoing studies as Language Assessment by Mechanical Turk,
using the abbreviation labMT 1.0 data set for the present work
(the full data set is provided as Supplementary Information for \cite{dodds2011e}).
Tabs.~\ref{tab:wordhap.happywords},
\ref{tab:wordhap.sadwords},
and
\ref{tab:wordhap.controversialwords}
respectively give the top 50 words according to positivity,
negativity, and standard deviation of happiness scores.

\section*{Results and Discussion}

\begin{figure*}[thp!]
  \centering
  \includegraphics[width=\textwidth]{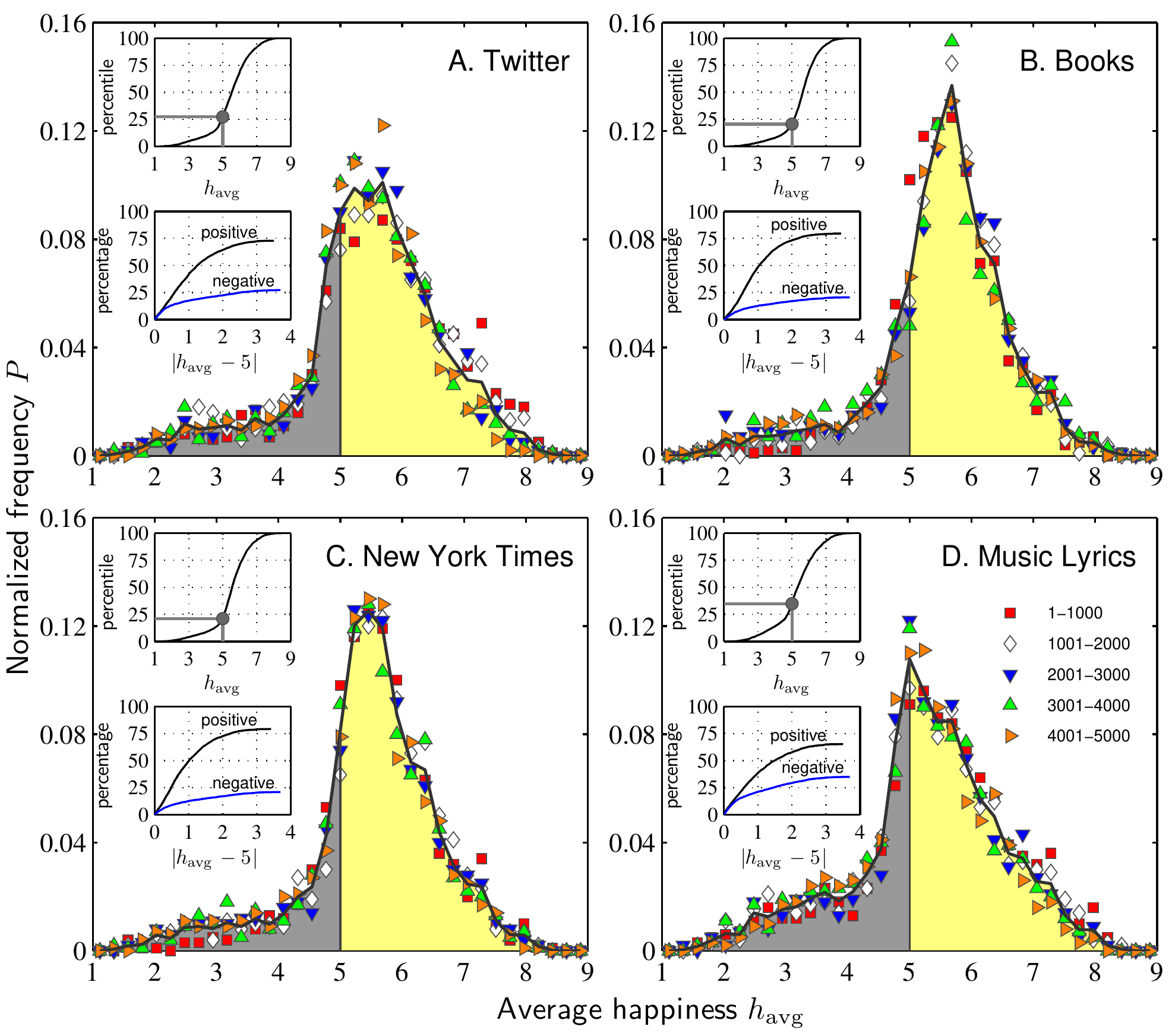}
  \caption{
    Positivity bias in the English language:
    normalized frequency distributions (solid black curves) of happiness scores
    for the 5000 most frequently used words in four corpora.
    Average happiness ratings for 10,222 words were obtained using Mechanical Turk 
    with 50 evaluations per word for a total of 501,110 human evaluations (see main text).
    The yellow shade indicates words with average happiness
    scores above the neutral value of 5, gray those below.
    The symbols show normalized frequency distributions 
    for words with given usage frequency ranks (see legend)
    suggesting a rough internal scale-free consistency of positivity
    Upper inset plots show percentile locations and the lower inset plots
    show the number of words found when cumulating toward the positive
    and negative sides of the neutral score of 5.
  }
  \label{fig:wordhap.dists}
\end{figure*}

\begin{figure*}[thp!]
  \centering
  \includegraphics[width=\textwidth]{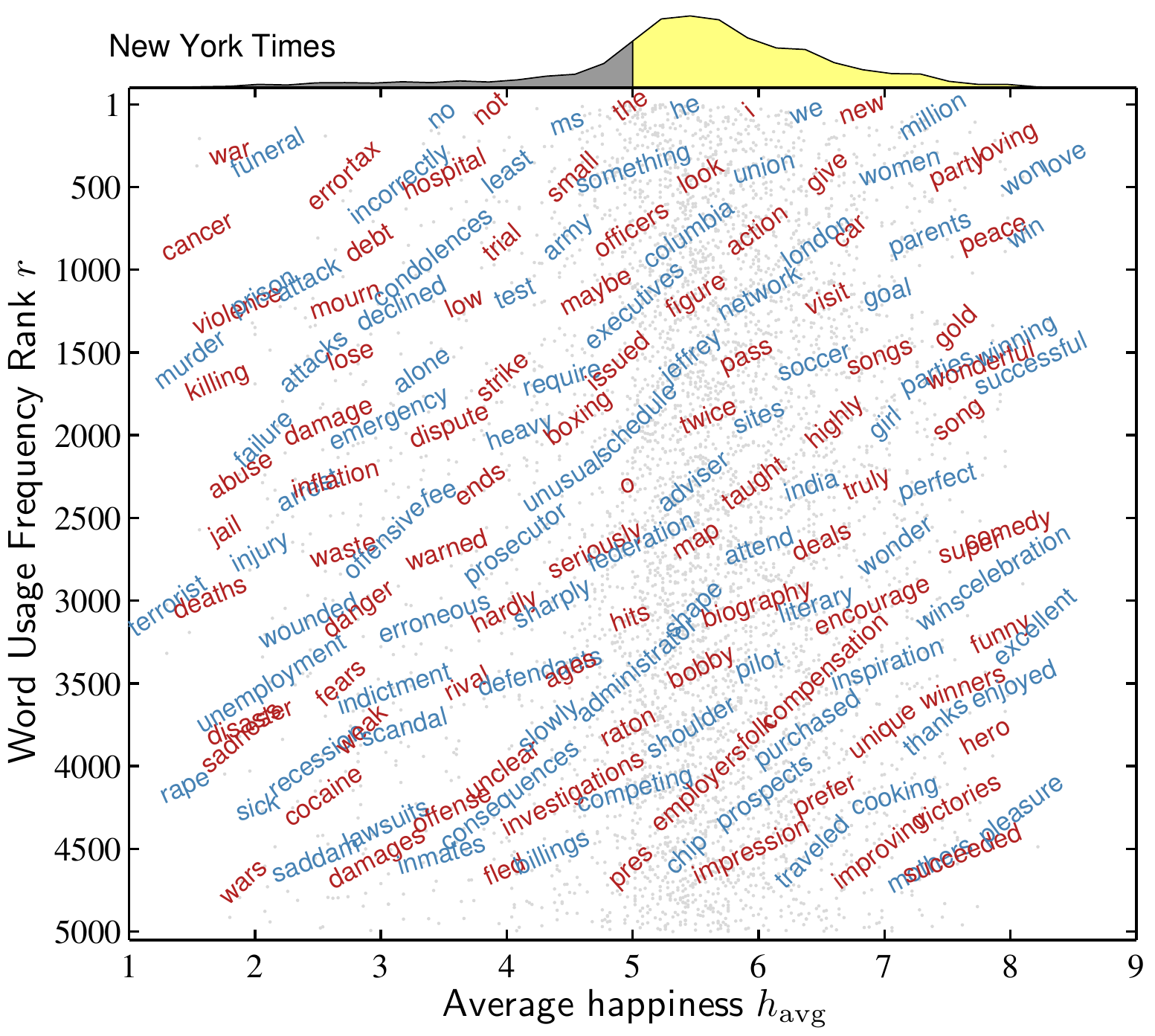}
  \caption{
    Example words for the New York Times as a function of 
    average happiness $h_{\rm avg}$ and
    usage frequency rank $r$.
    Words are centered at their values of $h_{\rm avg}$ and $r$,
    and angles and colors are only used
    for the purpose of readability.
    Each word is a representative of the set of words found
    in a rectangle of size 0.5 by 375 in $h_{\rm avg}$ and $r$,
    with all 5000 words located in the background by light gray points.
    The collapsed $h_{\rm avg}$ distribution at the top matches that shown
    in Fig.~\ref{fig:wordhap.dists}.
  }
  \label{fig:wordhap.words3}
\end{figure*}

\begin{figure*}[thp!]
  \centering
  \includegraphics[width=\textwidth]{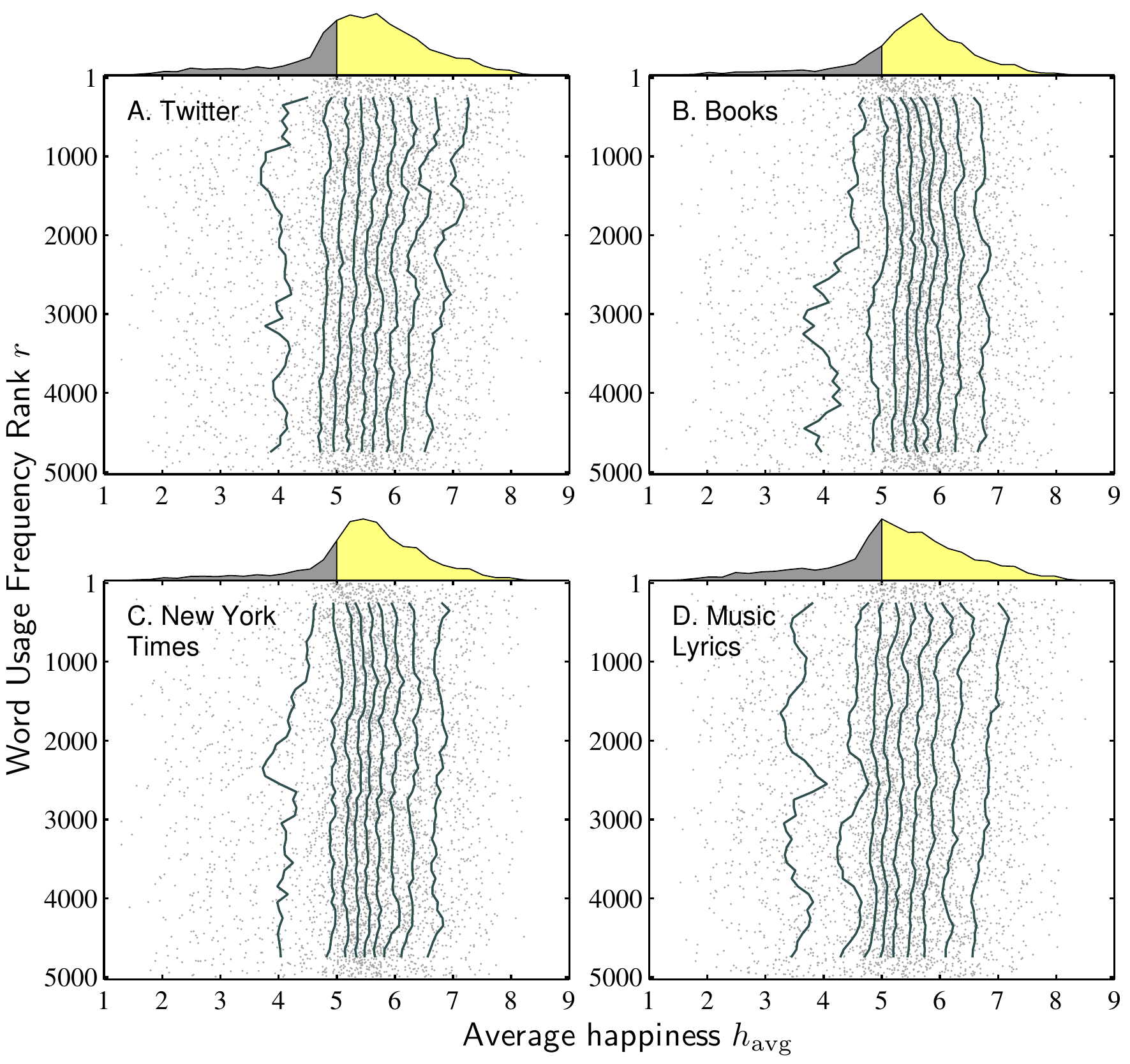}
  \caption{
    Deciles for average word happiness $h_{\rm avg}$ distributions as a function of 
    word usage frequency rank $r$.
    These `jellyfish plots' are created using a sliding window of 500 words
    moving down the vertical axis of usage frequency rank in increments of 100.
    The gray points mark $(h_{\rm avg}, r)$ for individual words, as in Fig.~\ref{fig:wordhap.words3}.
    The overall distributions of $h_{\rm avg}$, matching those
    in Fig.~\ref{fig:wordhap.dists}, cap each plot.
  }
  \label{fig:wordhap.percentiles}
\end{figure*}

\begin{figure*}[thp!]
  \centering
  \includegraphics[width=\textwidth]{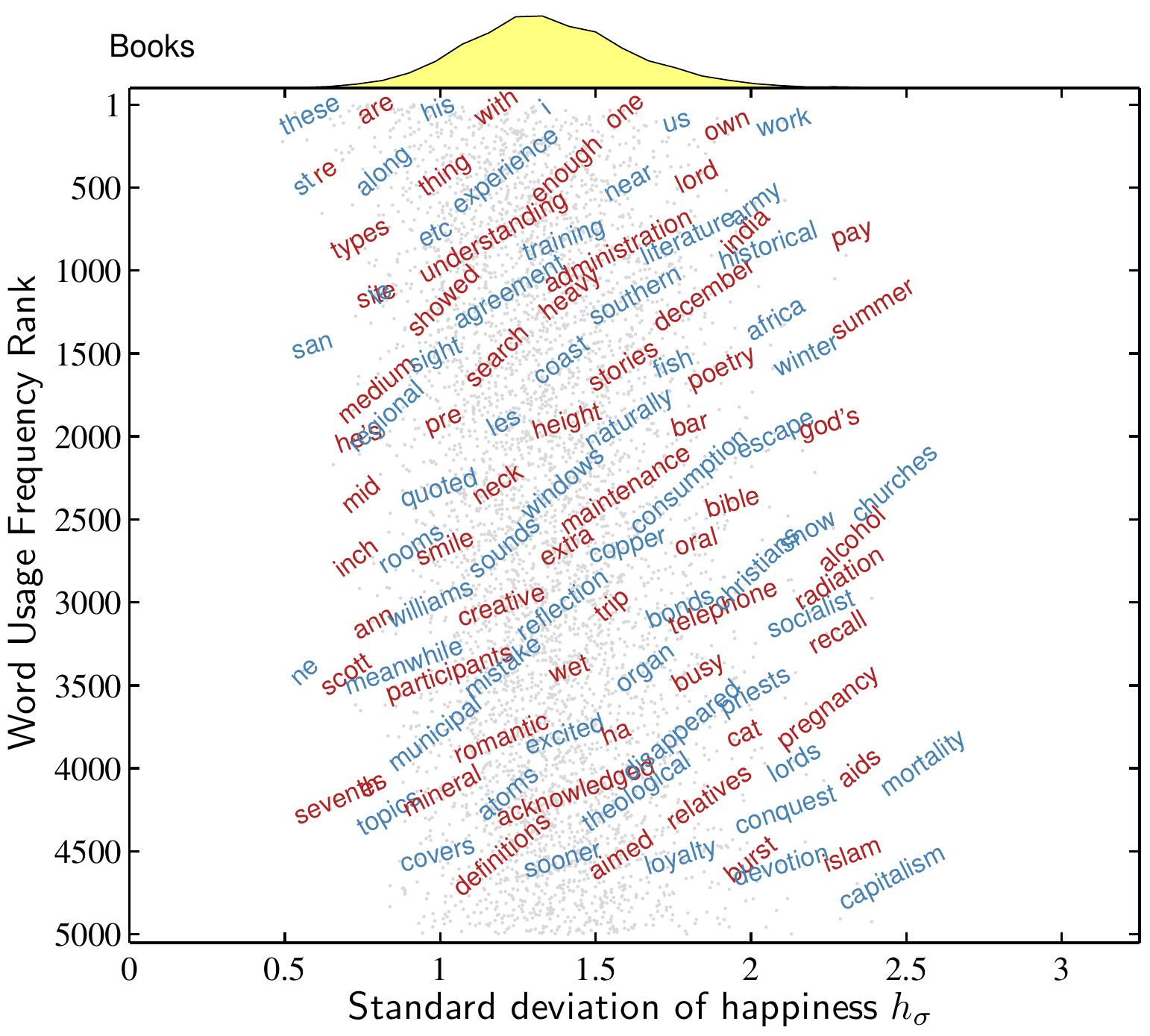}
  \caption{
    Example words for the Google Books corpus as a function of usage frequency rank
    and standard deviation of happiness estimates.
    Similar to Fig.~\ref{fig:wordhap.words3}, 
    each word shown represents all words in rectangles of size 0.2 and 375 in
    $h_{\sigma}$ and $r$.
    The histogram at the top of the figure represents
    the overall distribution for $h_{\sigma}$ for the
    first 5000 most frequent words.
    The light gray points indicate locations of the most frequent 5000 words in the Google Books corpus.
  }
  \label{fig:wordhap.stds2}
\end{figure*}

\begin{figure*}[thp!]
  \centering
  \includegraphics[width=\textwidth]{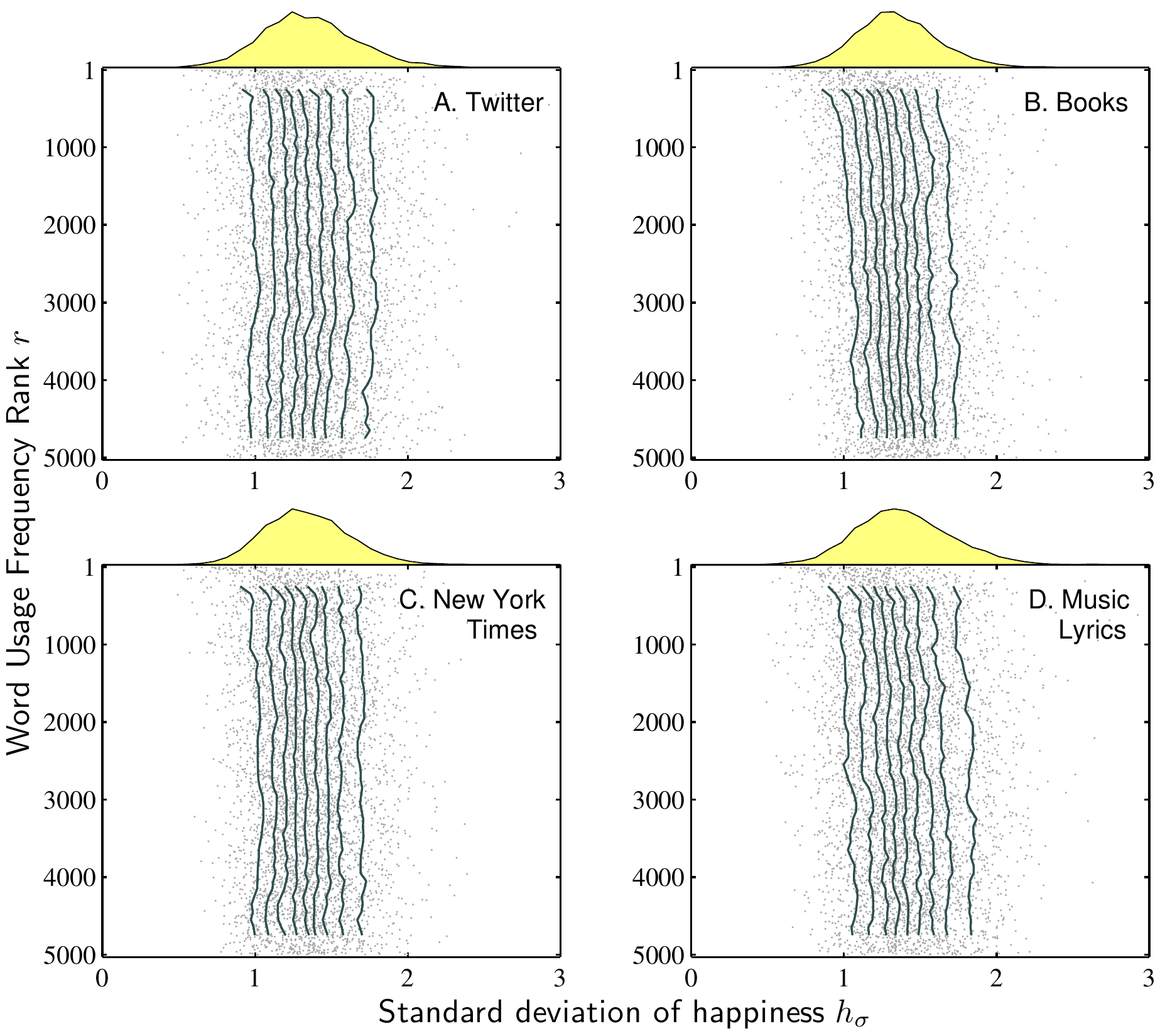}
  \caption{
    Deciles for standard deviations.
    As for Fig.~\ref{fig:wordhap.percentiles}, these `jellyfish plots'
    are created using a sliding window of 500 words
    moving across the horizontal axis of usage frequency rank increments of 100.
  }
  \label{fig:wordhap.stdspercentiles}
\end{figure*}

In Fig.~\ref{fig:wordhap.dists}, we show distributions
of average word happiness $h_{\rm avg}$ for our four corpora.
We first discuss the overall distributions,
i.e., those corresponding to the most frequent 5000 words combined
in each corpus (black curves), and then examine the robustness
of their forms with respect to frequency range.
The distributions as shown were formed using 35 equal-sized bins;
the number of bins does not change the visual form of the distributions
appreciably, and an odd number ensures that the neutral score of 5 is a bin center.
We employed binning only for visual display, 
using the raw data for all statistical analysis.

We see each distribution is unimodal and 
strongly positively skewed,
with a clear abundance of positive words 
($\havgfn > 5$, yellow shade) 
over negative ones
($\havgfn < 5$, gray shade).
In order, the percentages of positive words are
72.00\% (TW),
78.80\% (GB),
78.38\% (NYT),
and
64.14\% (ML).
Equivalently, and as further supported by 
Fig.~\ref{fig:wordhap.dists}'s upper inset plots
of percentile location,
we see the percentile corresponding
to the neutral score of 5 is well below
the median.
The lower inset plots show how the number of 
positive and negative words increase
as we cumulate moving away from the neutral
score of 5; positive words are always more abundant
further illustrating the positive bias.
The mode average happiness of words is 
either above neutral (TW, GB, and NYT)
or located there (ML).
Combining words across corpora, 
we also see the same overall positivity bias for parts of speech,
e.g., nouns and verbs (not shown),
in agreement with previous work~\cite{augustine2011a}.

While these overall distributions do not match in detail across corpora,
we do find they have an unexpected and striking internal consistency
with respect to usage frequency.  
We provide a series of increasingly refined and nuanced observations
regarding this emotional and linguistic phenomenon of scale invariance.

First, along with the overall distribution in each plot in Fig.~\ref{fig:wordhap.dists},
we also show distributions for subsets of 1000 words (symbols),
ordered by frequency rank $r$ (1--1000, 1001--2000, etc.).
The similarity of these distributions suggests 
to the eye that common and rare words are similarly distributed
in their perceived degree of positivity.

In Fig.~\ref{fig:wordhap.kstest}, we provide
statistical support via $p$-values from Kolmogorov-Smirnov tests for
each pairing of distributions.  
Here, $p$-values are to be interpreted as the probability
that two samples could have been derived from the same
underlying distribution.
The three corpora NYT, ML, and GB show the most
internal agreement, and
we see in all corpora that neighboring ranges of 1000 frequencies
could likely match in distribution.
Of the 40 pair-wise comparisons across the four corpora,
29 show statistically significant matches ($p > 10^{-2}$).

In any study of texts based on word counts, 
the words themselves need to be presented in some form
as commonsense checks on abstracted measurements.
To provide further insight into how word happiness 
behaves as a function of usage frequency rank, we plot a subsample of words
for the New York Times in Fig.~\ref{fig:wordhap.words3}.
We present analogous examples for the other three corpora 
in Figs.~\ref{fig:wordhap.words1},~\ref{fig:wordhap.words2}, 
and~\ref{fig:wordhap.words4}.
In these plots, usage frequency rank increases from bottom to top
with average happiness along the bottom axis.
To make clear the connection with Fig.~\ref{fig:wordhap.dists},
we include the overall distribution for the top 5000 words 
at the top of each plot.
Each word is centered at the location
of its values of $h_{\rm avg}$ and usage frequency rank.  
The alternating colors are used for visual clarity only, as are
the random angles.  
Underlying the words, the light gray points
indicate the locations of all of the most frequently used 5000 words.

For the New York Times example,
we find that the word pattern for average happiness and
usage frequency rank is indeed reasonable.
Down the right hand side of Fig.~\ref{fig:wordhap.words3},
we see highly positive words while decreasing in usage frequency such as 
`love', 
`win',
`comedy'
`celebration',
and
`pleasure'.
Similarly, down the left hand side, we find
`war',
`cancer',
`murder',
`terrorist',
and 
`rape'.
Words of flat affect such as 
`the',
`something',
`issued',
and
`administrator'
run down the middle of the happiness spectrum.
For words with usage frequency rank near 2500,
moving left to right in the plot,
we find the sequence of increasingly positive words
`jail',
`arrest',
`inflation',
`fee',
`ends',
`advisor',
`taught',
`india'
`truly',
and 
`perfect'.
Moving through the space represented in other directions
gives further reassurance of the general trends
we observe here.  Note that the random sampling 
of words used to generate these figures much more 
coarsely samples the word distributions for neutral
or medium levels of happiness.

While the four corpora share common words in their
most frequent 5000, numerous words appear in only one corpus.
For example, `rainbows' and `kissing' make the top 5000 only
for Music Lyrics, and `punishment' the same for the Google Books corpus
(see Tabs.~\ref{tab:wordhap.happywords} and~\ref{tab:wordhap.sadwords}).
Moreover, the usage frequency rankings change strongly,
as a visual comparison of 
Fig.~\ref{fig:wordhap.words3}
with 
Figs.~\ref{fig:wordhap.words1},~\ref{fig:wordhap.words2}, 
and~\ref{fig:wordhap.words4} reveals.
Further detailed comparisons can be made directly from
the labMT 1.0 data set~\cite{dodds2011e}.

To bolster our observations quantitatively,
we first compute a linear regression
and a Spearman correlation coefficient $\rho_s$
and associated $p$-value (two-sided) for $h_{\rm avg}$ as a function of usage frequency rank, $r$.
We record the results for each corpus in Tab.~\ref{tab:wordhap.havgcorr}.

The slopes of linear fits are all negative but
extremely small, ranging from 
-3.04$\times$$10^{-5}$ (GB)
to
-7.78$\times$$10^{-5}$ (TW).
All corpora also present a weak negative correlation, ranging from $\rho_{\rm s}=-0.013$ (GB)
to -0.103 (TW).  
The correlation for
the Google Books corpus is not statistically significant ($p$=0.35), while
it is for the other three, and especially so for TW and ML
($p$ = 2.3$\times$$10^{-13}$ and 1.0$\times$$10^{-8}$).

\begin{table}[htbp!]
  \centering
  \begin{tabular}{|l|c|c|c|c|}
    \hline
    \hline
    Corpus & $\alpha$ & $\beta$ &  $\rho_{\rm s}$ & $p$-value \\
    \hline
    Twitter & -7.78$\times$$10^{-5}$ & 5.67 & -0.103 & 2.3$\times$$10^{-13}$ \\
    Books  & -3.04$\times$$10^{-5}$ & 5.62 & -0.013 & 3.5$\times$$10^{-1}$ \\
    New York Times & -4.17$\times$$10^{-5}$ & 5.61 &  -0.0437 & 2.0$\times$$10^{-3}$ \\
    Music Lyrics: & -6.12$\times$$10^{-5}$ & 5.45 & -0.0808 & 1.0$\times$$10^{-8}$ \\
    \hline
    \hline
  \end{tabular}
  \caption{
    Linear fit coefficients, Spearman correlation coefficients, and $p$-values 
    for average word happiness $h_{\rm avg}$
    as a function of usage frequency rank $r$.
    Fit is $h_{\rm avg} = \alpha r + \beta$.
  }
  \label{tab:wordhap.havgcorr}
\end{table}

We next move to a more detailed quantitative view of 
the word happiness distribution
as a function of word usage frequency.
In Fig.~\ref{fig:wordhap.percentiles},
we show how deciles behave as a function of usage frequency rank.
Using a sliding window containing 500 words, 
we compute deciles moving down the usage frequency rank axis.
Using these `jellyfish plots',
we see that apart from the lowest decile (which
is universally uneven), 
GB and NYT are very stable while a slight negative
trend is perceptible for TW and ML.
We can now with some confidence state
that the measured, edited writing of the New York Times
and the Google Books corpus possess a 
remarkable scale invariance
in emotion with respect to word usage frequency.
The emotional content of words on Twitter and in music lyrics, while
still roughly similar across usage frequency ranks, 
show a small bias 
towards common words being disproportionately positive in 
comparison with increasing rare ones.
The bias is sufficiently small as to
be likely indiscernible by an individual familiar
with these corpora; moreover, cognitive biases regarding
the salience of information would presumably render
such detection impossible~\cite{baumeister2001a}.

We have thus far considered distributions of
average happiness values for words.
Each word's estimate comes from a distribution
of assessment scores, and a useful, simple investigation
can be carried out on the standard deviation of
individual word happiness, $h_{\sigma}$.

A range of word and concept categories 
yielded high $h_{\sigma}$ in our study, the top 50 of 
which are shown in Tab.~\ref{tab:wordhap.controversialwords}.
At the top of the list, we observe words
that are or relate to profanities, alcohol and tobacco, religion, 
both capitalism and socialism, sex, marriage, 
fast foods, climate, and cultural
phenomena such as the Beatles, the iPhone, and zombies.
As a result of variation in the rater's preferences 
perhaps due to inherent controversy or cultural and demographic variation, 
these terms all elicited diverse responses.

We repeat our analyses of $h_{\rm avg}$ for $h_{\sigma}$
by first considering a sample of words for the Google Books corpus,
Fig~\ref{fig:wordhap.stds2},
and then the behavior of deciles, Fig.~\ref{fig:wordhap.stdspercentiles}.
(In Fig~\ref{fig:wordhap.dists-stds} we present
the overall distributions, the equivalent of Fig.~\ref{fig:wordhap.dists}.)
For our entire collection of words, we find most
values of $h_{\sigma}$ fall in the range $[0.5, 2.5]$.

In Fig.~\ref{fig:wordhap.stds2}, 
we show example words from the Google Books corpus as a 
function of word usage frequency rank and standard deviation 
(Figs.~\ref{fig:wordhap.stds1}, \ref{fig:wordhap.stds3},
and \ref{fig:wordhap.stds4} show the same for TW, NYT, and ML
).
The right hand side of Fig.~\ref{fig:wordhap.stds2}
shows example words with high $h_{\sigma}$ and 
increasing usage frequency rank
including `work',
`pay',
`summer',
`churches',
`mortality'
and 
`capitalism'.
For low $h_{\sigma}$
(the left hand side of Fig.~\ref{fig:wordhap.stds2}),
we see basic, neutral words such as
`these',
`types',
`inch',
and 
`seventh'.

While this word diagram is primarily intended for qualitative purposes, 
we see that for $h_{\sigma}$,
the overall trend for Google Books 
is a gradual increase as a function of usage frequency rank.
In other words, relatively rarer words have higher standard deviations
in comparison with relatively more common ones.
This is confirmed visually in Fig.~\ref{fig:wordhap.stdspercentiles}, 
where we present jellyfish plots showing deciles for all four corpora.  
The Music Lyrics
corpus shows a similar increase in $h_{\sigma}$ with usage frequency rank as GB,
whereas TW and NYT corpora exhibit no obvious linear variation.
These observations are supported by the linear fits
and Spearman correlation coefficients recorded in Tab.~\ref{tab:wordhap.hstdcorr},
where we consider $h_{\sigma}$ as a function of usage frequency rank.
All linear approximations yield 
a very small positive growth, with both the TW and NYT corpora
clearly smaller than the other two, particularly TW.
The corresponding Spearman correlation coefficients indicate
we have statistically significant monotonic growth in $h_{\sigma}$ for
GB, ML, and NYT, particularly the first two, and indicates no evidence
of growth for TW.  

All told, we find slight deviation from an exact scaling independence of $h_{\rm avg}$ and
$h_{\sigma}$ in terms of usage frequency rank, but it is highly constrained and
corpus specific. 
In particular, the corpora that show a slight negative correlation between $h_{\rm avg}$ and usage frequency rank,
TW and ML, do not match those showing a positive correlation between
$h_{\sigma}$ and usage frequency rank, GB and ML.

\begin{table}[htbp!]
  \centering
  \begin{tabular}{|l|c|c|c|c|}
    \hline
    \hline
    Corpus & $\alpha$ & $\beta$ & $\rho_{\rm s}$ & $p$-value \\
    \hline
    Twitter & 1.47$\times$$10^{-6}$ & 1.35 & 0.0116 & 4.1$\times$$10^{-1}$ \\
    Books   & 3.36$\times$$10^{-5}$ & 1.27 & 0.176  & 5.0$\times$$10^{-36}$ \\
    New York Times & 9.33$\times$$10^{-6}$ & 1.32 & 0.0439 & 1.9$\times$$10^{-3}$ \\
    Music Lyrics & 2.76$\times$$10^{-5}$ & 1.33 & 0.134 & 1.6$\times$$10^{-21}$ \\
    \hline
    \hline
  \end{tabular}
  \caption{
    Spearman correlation coefficients for standard
    deviation of word happiness estimates
    as a function of usage frequency rank.
    Fit is $h_{\sigma} = \alpha r + \beta$.
  }
  \label{tab:wordhap.hstdcorr}
\end{table}

\section*{Concluding remarks}

Our findings are that positive words strongly outnumber negative words overall,
and that there is a very limited, corpus-specific tendency for high frequency words to be more
positive than low frequency words.
These two aspects of positivity and usage frequency can
only be separated with the kind of data we study here.
Previous claims that positive words are used more frequently~\cite{boucher1969a,rozin2010a,augustine2011a},
suffered from insufficient, non-representative data.
For example, Rozin et al.\ recently compared usage 
frequencies for just seven adjective pairs of positive-negative opposites~\cite{rozin2010a}.
Augustine et al.\ showed that average happiness
and usage frequencies for 1034 words~\cite{bradley1999a}
were more positively correlated than we observe here~\cite{augustine2011a}; 
however, since these words were chosen
for their meaningful nature~\cite{bradley1999a,mehrabian1974a,bellezza1986a}
rather than by their rate of occurrence, 
their findings are naturally tempered.
A positivity bias is also not inconsistent with many
observations that negative emotions in isolation 
are more potent and diverse than positive words~\cite{baumeister2001a}.

In sum, our findings for these diverse English language corpora
suggest that a positivity bias is universal, that the emotional
spectrum of language is very close to self-similar with respect to frequency,
and that in our stories and writings we tend toward prosocial communication.
Our work calls for similar studies of other languages
and dialects, 
examinations of corpora factoring in popularity (e.g., of books or articles),
as well as investigations
of other more specific emotional dimensions.
Related work would explore changes in positivity bias over time,
and correlations with quantifiable 
aspects of societal organization and function
such as wealth, cultural norms, and political structures.
Analyses of the emotional content of 
phrases and sentences in large-scale texts would also be a 
natural next, more complicated stage of research.
Promisingly, we have shown elsewhere for Twitter
that the average happiness of individual words correlates well with 
that of surrounding words in status updates~\cite{dodds2011e}.

The authors are indebted to conversations
with 
B.~Tivnan,
N.~Johnson,
and
A.~Reece.
\revtexonly{The authors are grateful for the computational resources provided by the Vermont Advanced Computing Center which is supported by NASA (NNX 08A096G). KDH was supported by VT-NASA EPSCoR. PSD was supported by NSF CAREER Award \# 0846668.}

\clearpage

\setcounter{page}{1}
\renewcommand{\thepage}{S\arabic{page}}
\renewcommand{\thefigure}{S\arabic{figure}}
\renewcommand{\thetable}{S\arabic{table}}
\setcounter{figure}{0}
\setcounter{table}{0}

{\large\textbf{Supplementary Information}}\begin{figure*}[thp!]  
  \centering
  \includegraphics[width=\textwidth]{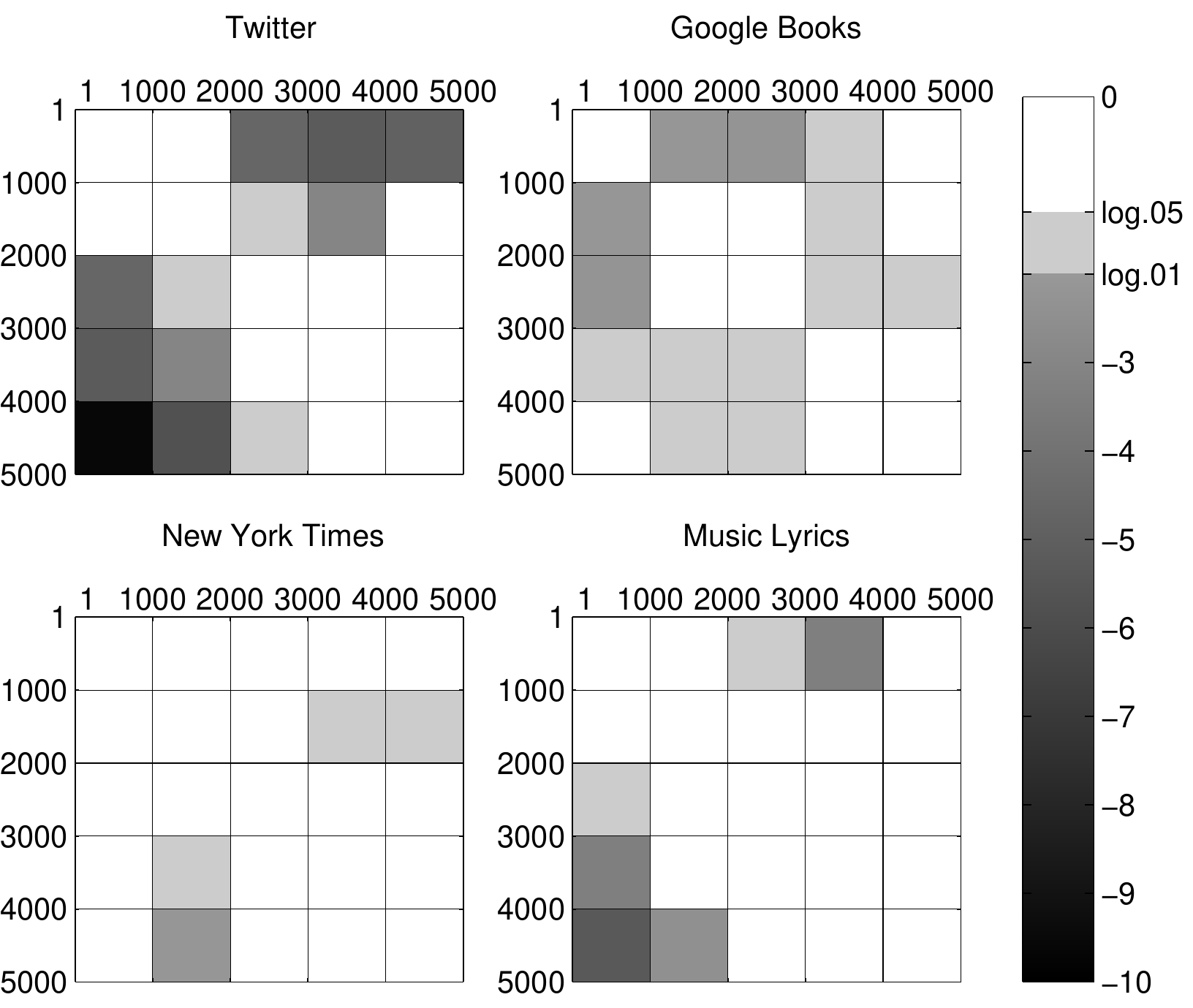}
  \caption{
    Results of Kolmogorov-Smirnov tests comparing word happiness
    distributions shown in Fig.~\ref{fig:wordhap.dists}.
    For each corpus, the $p$-value reports the probability
    that the two samples being compared could come from
    the same distribution with lighter colors meaning more likely.
    The gray-scale corresponds to $\log_{10} (p\mbox{-value})$.
    }
  \label{fig:wordhap.kstest}
\end{figure*}

\begin{figure*}[thp!]
  \centering
  \includegraphics[width=\textwidth]{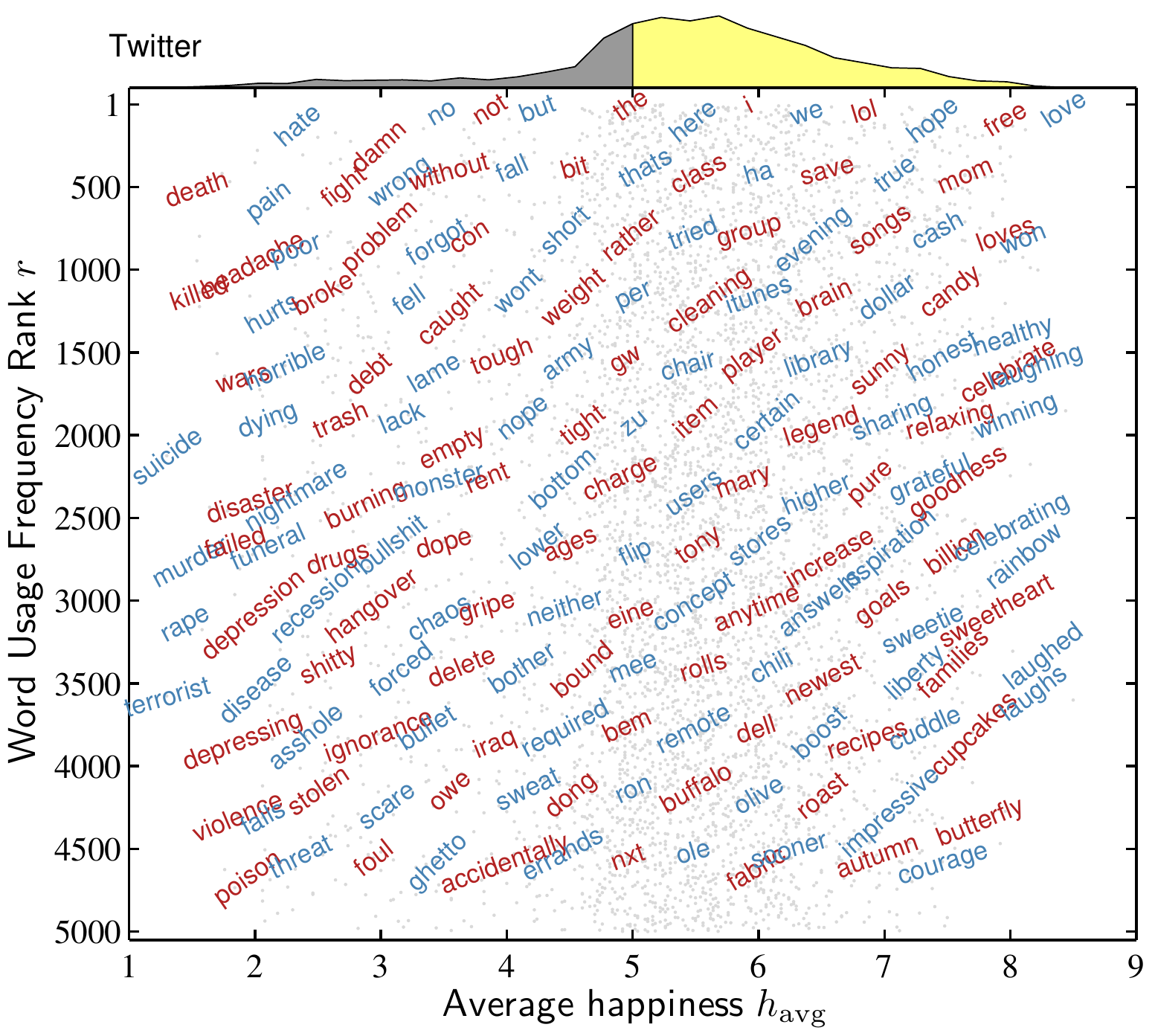}
  \caption{
    Example words for Twitter as a function of usage frequency rank
    and average happiness.
  }
  \label{fig:wordhap.words1}
\end{figure*}

\begin{figure*}[thp!]
  \centering
  \includegraphics[width=\textwidth]{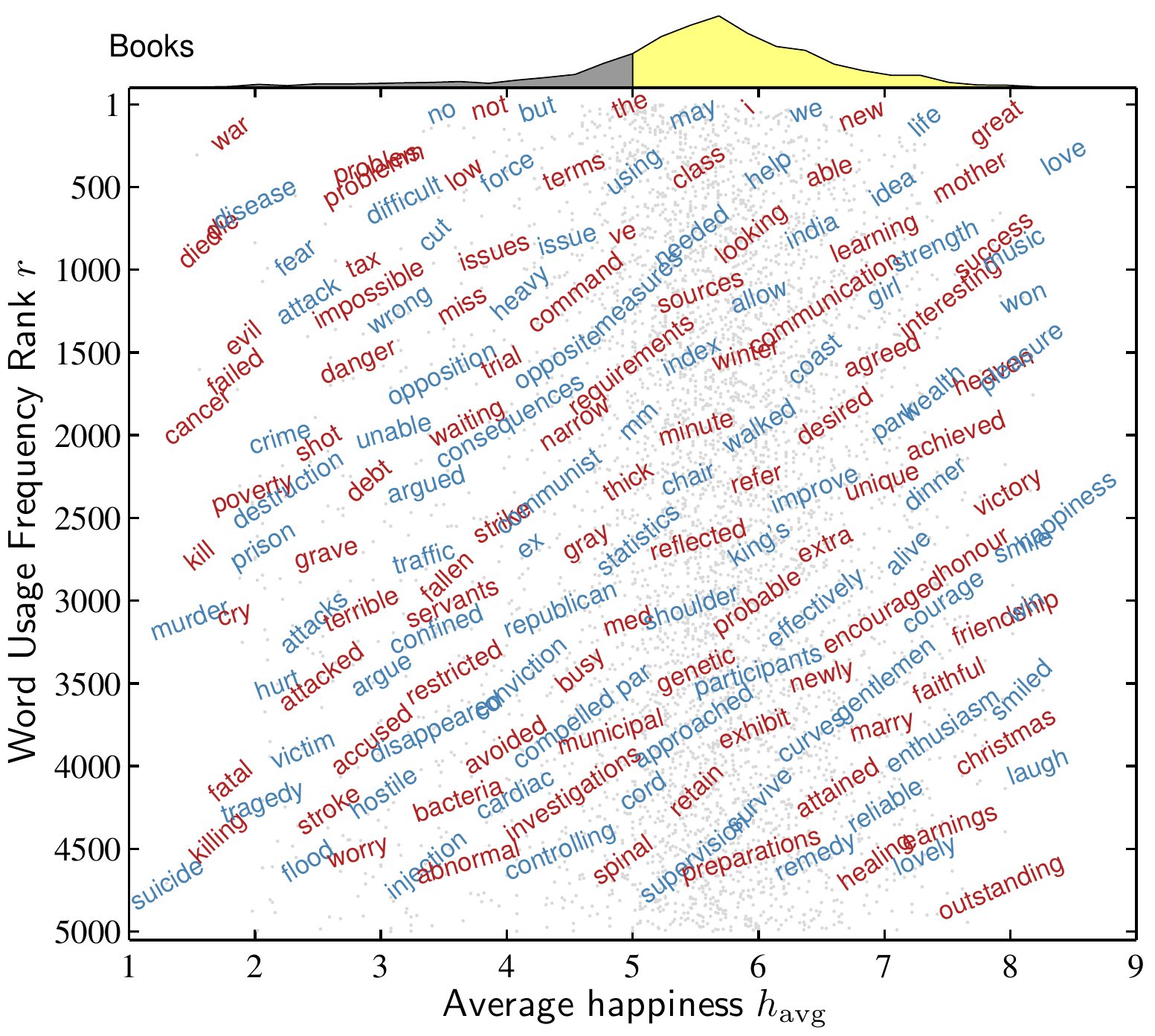}
  \caption{
    Example words for the Google Books corpus as a function of usage frequency rank
    and average happiness.
  }
  \label{fig:wordhap.words2}
\end{figure*}

\begin{figure*}[thp!]
  \centering
  \includegraphics[width=\textwidth]{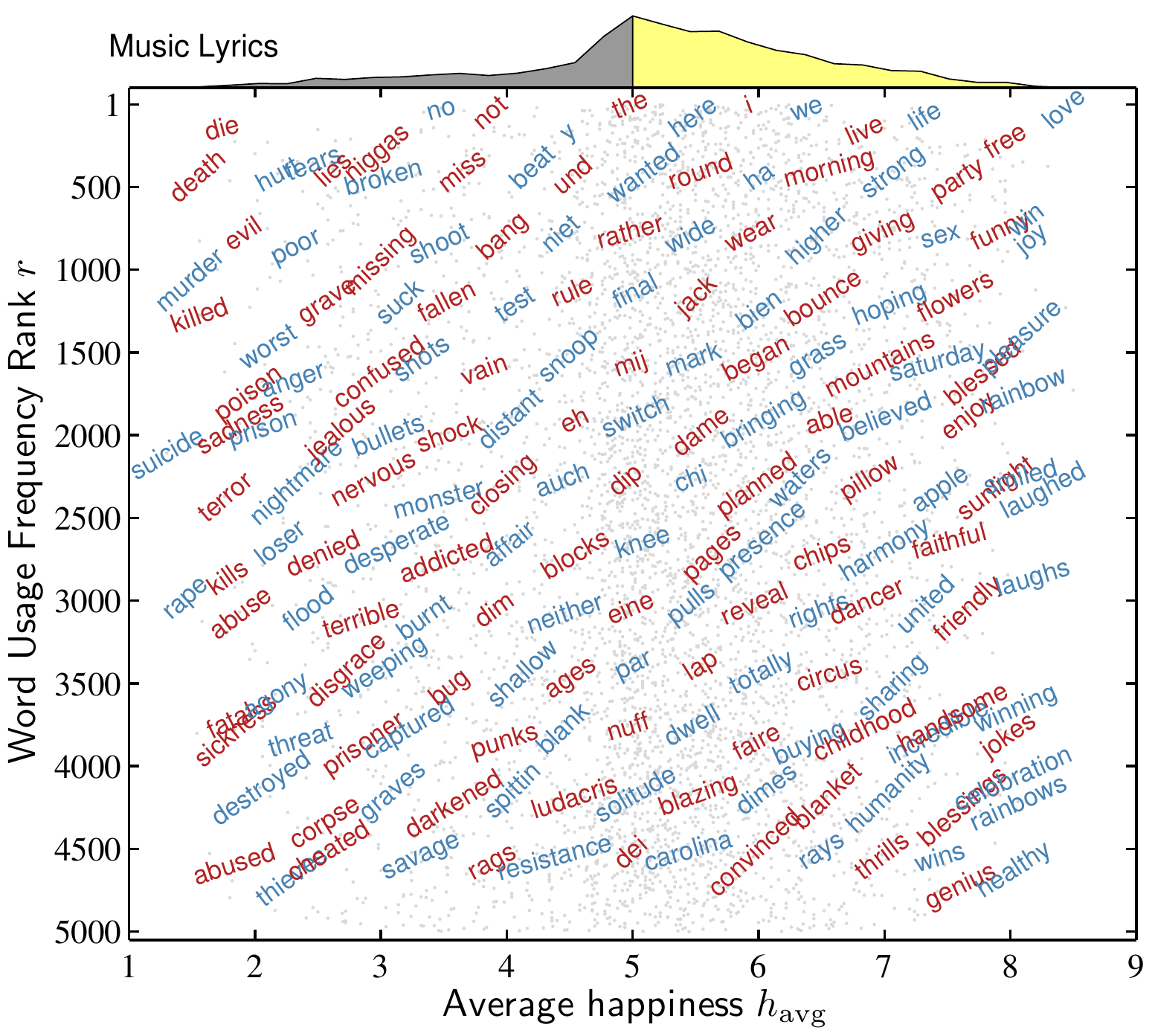}
  \caption{
    Example words for the Music Lyrics corpus as a function of usage frequency rank
    and average happiness.
  }
  \label{fig:wordhap.words4}
\end{figure*}

\begin{figure*}[thp!]
  \centering
  \includegraphics[width=\textwidth]{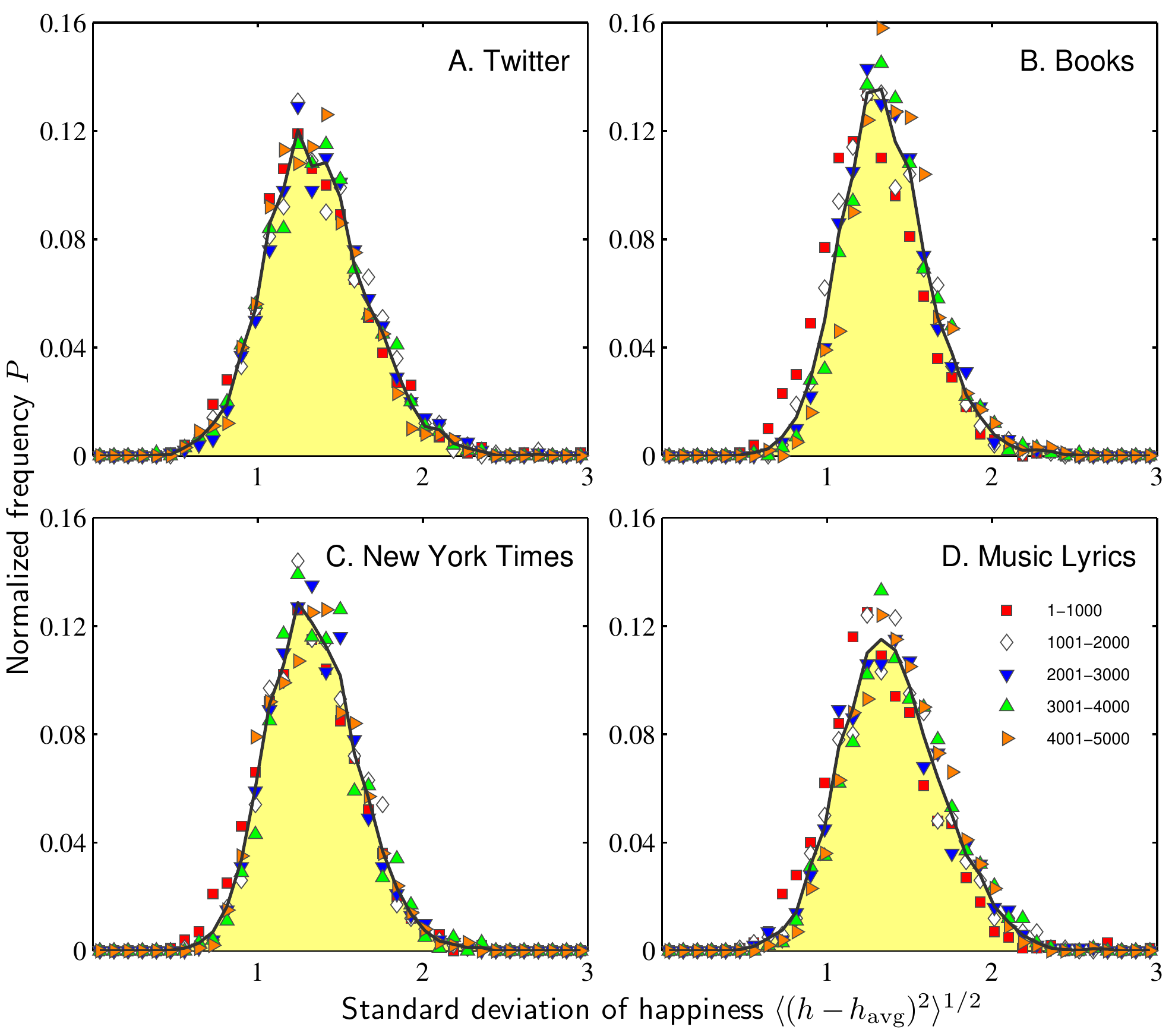}
  \caption{
    Overall distributions of standard deviations in happiness scores
    for the four corpora.  As with average happiness, distributions
    for subsets of usage frequency ranks (symbols, see legend) 
  }
  \label{fig:wordhap.dists-stds}
\end{figure*}

\begin{figure*}[thp!]
  \centering
  \includegraphics[width=\textwidth]{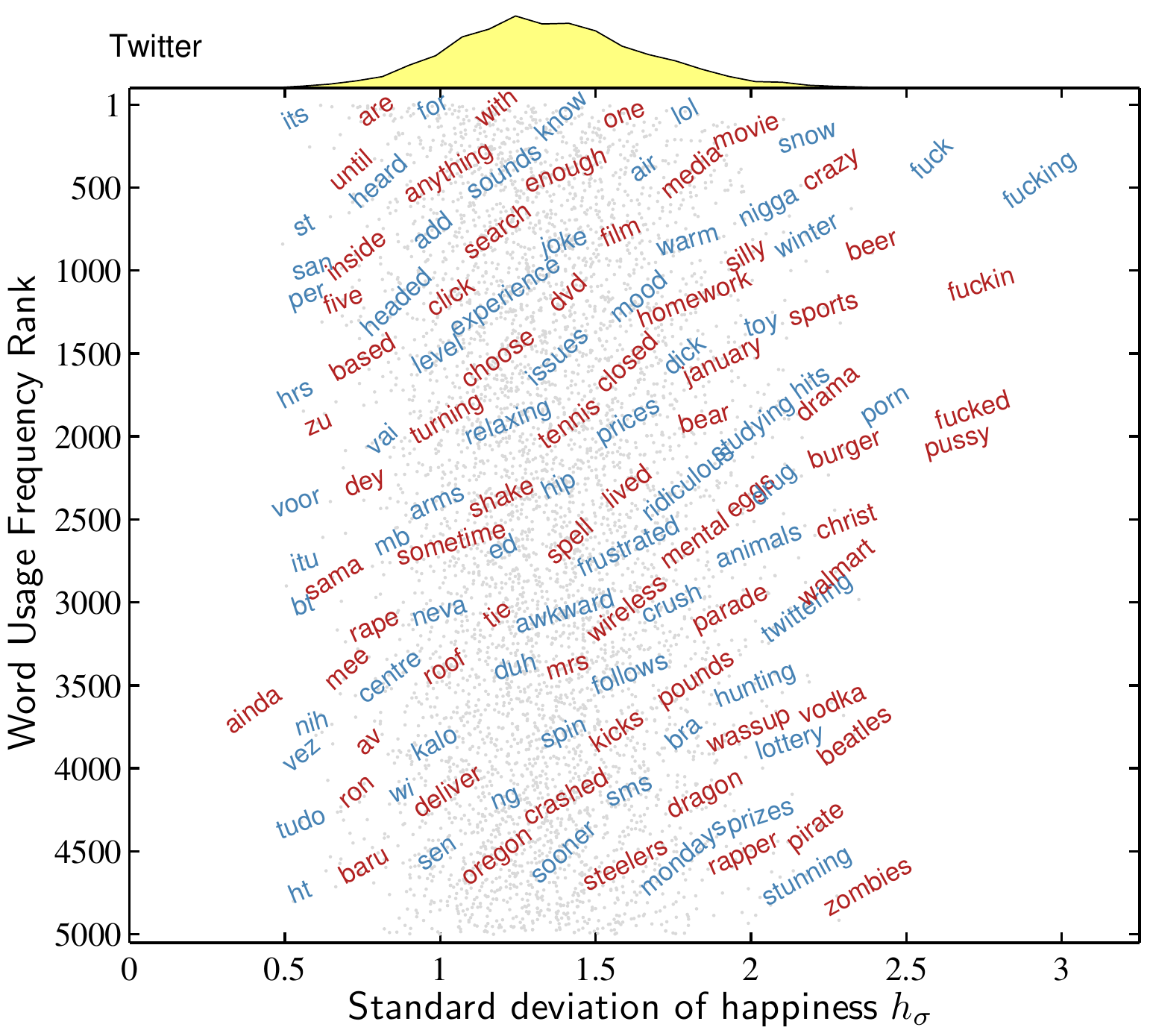}
  \caption{
    Example words for Twitter as a function of usage frequency rank
    and standard deviation of happiness estimates.
  }
  \label{fig:wordhap.stds1}
\end{figure*}

\begin{figure*}[thp!]
  \centering
  \includegraphics[width=\textwidth]{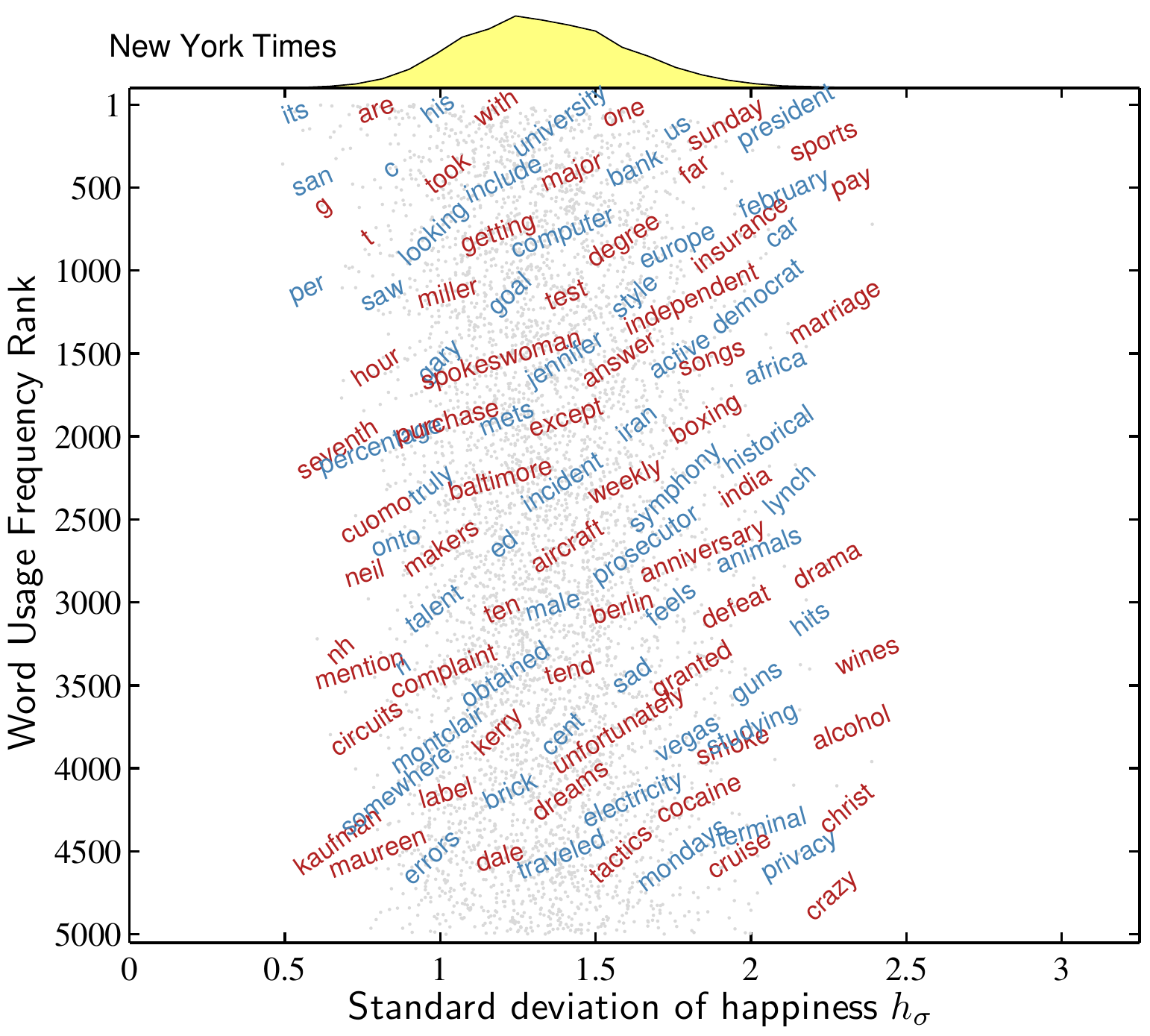}
  \caption{
    Example words for the New York Times as a function of usage frequency rank
    and standard deviation of happiness estimates.
  }
  \label{fig:wordhap.stds3}
\end{figure*}

\begin{figure*}[thp!]
  \centering
  \includegraphics[width=\textwidth]{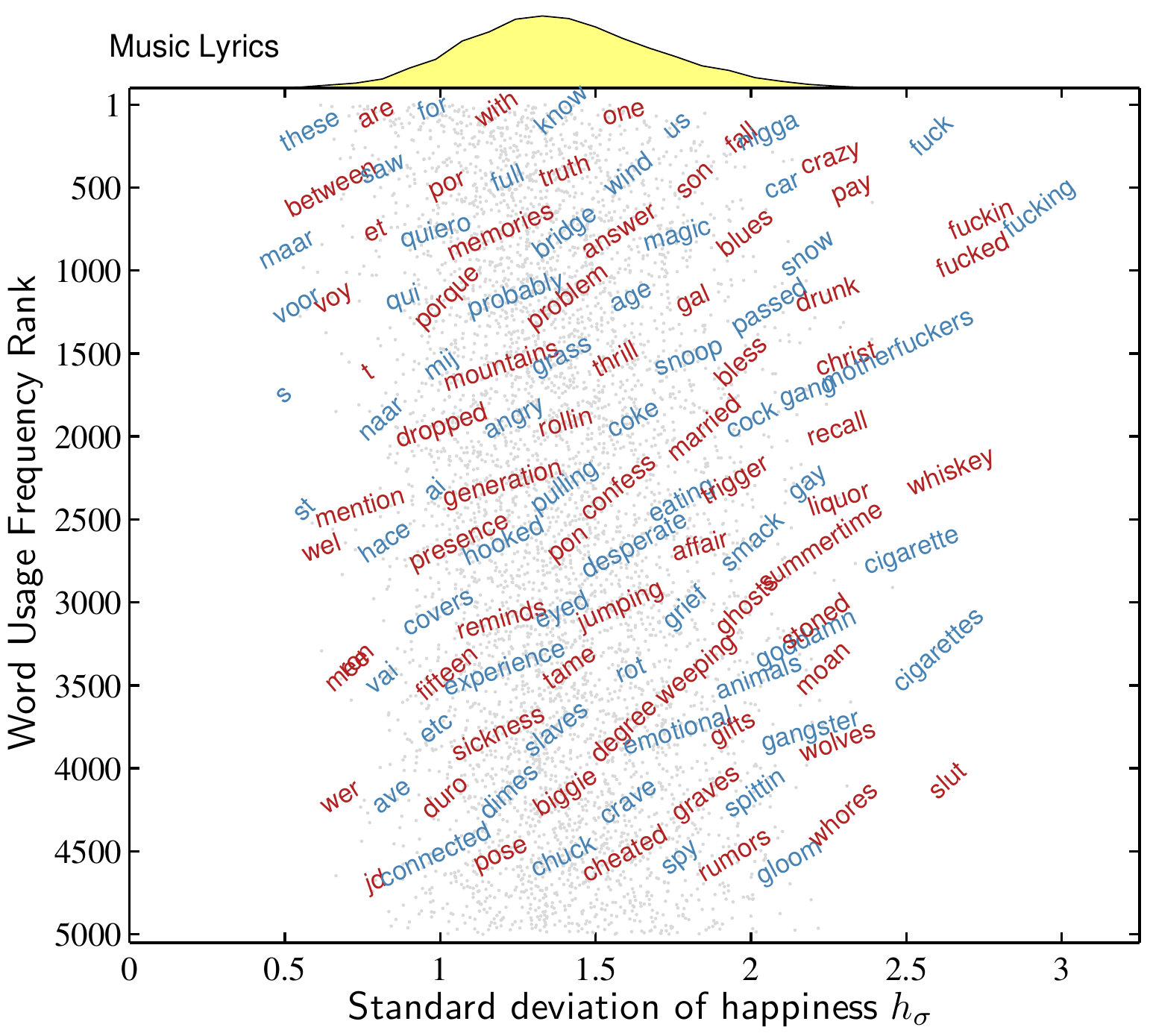}
  \caption{
    Example words for the Music Lyrics corpus as a function of usage frequency rank
    and standard deviation of happiness estimates.
  }
  \label{fig:wordhap.stds4}
\end{figure*}

\begin{table*}[htbp]
  \centering
  \begin{tabular}{rrrrrrrr}
    \hline
    $h_{\rm rank}$ & word  & $h_{\rm avg}$ & $h_{\sigma}$ & TW rank & GB rank & NYT rank  & ML rank \\
    \hline
    \hline
    1       & laughter        & 8.50    & 0.9313  & 3600    & --      & --      & 1728 \\
    2       & happiness       & 8.44    & 0.9723  & 1853    & 2458    & --      & 1230 \\
    3       & love    & 8.42    & 1.1082  & 25      & 317     & 328     & 23 \\
    4       & happy   & 8.30    & 0.9949  & 65      & 1372    & 1313    & 375 \\
    5       & laughed & 8.26    & 1.1572  & 3334    & 3542    & --      & 2332 \\
    6       & laugh   & 8.22    & 1.3746  & 1002    & 3998    & 4488    & 647 \\
    7       & laughing        & 8.20    & 1.1066  & 1579    & --      & --      & 1122 \\
    8       & excellent       & 8.18    & 1.1008  & 1496    & 1756    & 3155    & -- \\
    9       & laughs  & 8.18    & 1.1551  & 3554    & --      & --      & 2856 \\
    10      & joy     & 8.16    & 1.0568  & 988     & 2336    & 2723    & 809 \\
    11      & successful      & 8.16    & 1.0759  & 2176    & 1198    & 1565    & -- \\
    12      & win     & 8.12    & 1.0812  & 154     & 3031    & 776     & 694 \\
    13      & rainbow & 8.10    & 0.9949  & 2726    & --      & --      & 1723 \\
    14      & smile   & 8.10    & 1.0152  & 925     & 2666    & 2898    & 349 \\
    15      & won     & 8.10    & 1.2164  & 810     & 1167    & 439     & 1493 \\
    16      & pleasure        & 8.08    & 0.9655  & 1497    & 1526    & 4253    & 1398 \\
    17      & smiled  & 8.08    & 1.0660  & --      & 3537    & --      & 2248 \\
    18      & rainbows        & 8.06    & 1.3603  & --      & --      & --      & 4216 \\
    19      & winning & 8.04    & 1.0490  & 1876    & --      & 1426    & 3646 \\
    20      & celebration     & 8.02    & 1.5318  & 3306    & --      & 2762    & 4070 \\
    21      & enjoyed & 8.02    & 1.5318  & 1530    & 2908    & 3502    & -- \\
    22      & healthy & 8.02    & 1.0593  & 1393    & 3200    & 3292    & 4619 \\
    23      & music   & 8.02    & 1.1156  & 132     & 875     & 167     & 374 \\
    24      & celebrating     & 8.00    & 1.1429  & 2550    & --      & --      & -- \\
    25      & congratulations & 8.00    & 1.6288  & 2246    & --      & --      & -- \\
    26      & weekend & 8.00    & 1.2936  & 317     & --      & 833     & 2256 \\
    27      & celebrate       & 7.98    & 1.1516  & 1606    & --      & 3574    & 2108 \\
    28      & comedy  & 7.98    & 1.1516  & 1444    & --      & 2566    & -- \\
    29      & jokes   & 7.98    & 0.9792  & 2812    & --      & --      & 3808 \\
    30      & rich    & 7.98    & 1.3169  & 1625    & 1221    & 1469    & 890 \\
    31      & victory & 7.98    & 1.0784  & 1809    & 2341    & 687     & 2845 \\
    32      & christmas       & 7.96    & 1.2930  & 138     & 3846    & 2097    & 599 \\
    33      & free    & 7.96    & 1.2610  & 85      & 342     & 393     & 219 \\
    34      & friendship      & 7.96    & 1.1241  & 4273    & 3098    & 3669    & 3980 \\
    35      & fun     & 7.96    & 1.3087  & 110     & 4135    & 2189    & 463 \\
    36      & holidays        & 7.96    & 1.2610  & 1204    & --      & --      & -- \\
    37      & loved   & 7.96    & 1.1599  & 465     & 2178    & 890     & 517 \\
    38      & loves   & 7.96    & 1.3696  & 780     & --      & --      & 653 \\
    39      & loving  & 7.96    & 1.0093  & 947     & 4396    & 230     & 527 \\
    40      & beach   & 7.94    & 1.0577  & 573     & 3596    & 551     & 1475 \\
    41      & hahaha  & 7.94    & 1.5572  & 428     & --      & --      & -- \\
    42      & kissing & 7.94    & 1.1323  & --      & --      & --      & 2052 \\
    43      & sunshine        & 7.94    & 1.1678  & 2080    & --      & --      & 950 \\
    44      & beautiful       & 7.92    & 1.1753  & 266     & 1159    & 1754    & 467 \\
    45      & delicious       & 7.92    & 1.2591  & 1565    & --      & --      & -- \\
    46      & friends & 7.92    & 1.1925  & 258     & 658     & 347     & 321 \\
    47      & funny   & 7.92    & 1.0467  & 358     & --      & 3194    & 755 \\
    48      & outstanding     & 7.92    & 1.1400  & 4468    & 4721    & 1797    & -- \\
    49      & paradise        & 7.92    & 1.3974  & 3096    & --      & --      & 1146 \\
    50      & sweetest        & 7.92    & 1.2911  & --      & --      & --      & 2232 \\
    \hline
  \end{tabular}
  \caption{
    The 50 most positive words, as assessed by our Mechanical Turk survey.
    Rankings of each word in the four corpora are provided.
    A `--' indicates a word was not in the most frequent 5000 words in the given corpus.
  }
  \label{tab:wordhap.happywords}
\end{table*}

\begin{table*}[htbp]
  \centering
  \begin{tabular}{rrrrrrrr}
    \hline
    $h_{\rm rank}$ & word  & $h_{\rm avg}$ & $h_{\sigma}$ & TW rank & GB rank & NYT rank  & ML rank \\
    \hline
    \hline
    10173   & disease       & 2.00  & 1.3093        & 3531  & 598   & 1391  & 1780 \\
    10174   & illness       & 2.00  & 1.1780        & --    & 2738  & 1690  & -- \\
    10175   & killers       & 2.00  & 1.5253        & --    & --    & --    & 3303 \\
    10176   & punishment    & 2.00  & 1.3401        & --    & 2750  & --    & -- \\
    10177   & criminal      & 1.98  & 1.2696        & 2722  & 2421  & 1322  & 3261 \\
    10178   & depression    & 1.98  & 1.5583        & 3082  & 2406  & --    & -- \\
    10179   & headache      & 1.98  & 1.1156        & 959   & --    & --    & -- \\
    10180   & poverty       & 1.98  & 1.1156        & --    & 2343  & 3744  & -- \\
    10181   & tumors        & 1.98  & 1.3461        & --    & 4876  & --    & -- \\
    10182   & bomb  & 1.96  & 1.2771        & 1292  & --    & 2815  & 1227 \\
    10183   & disaster      & 1.96  & 1.4280        & 2399  & --    & 3729  & 3355 \\
    10184   & fail  & 1.96  & 1.0294        & 1160  & 2481  & 4030  & 1758 \\
    10185   & poison        & 1.94  & 1.1502        & 4668  & --    & --    & 1740 \\
    10186   & depressing    & 1.90  & 1.2164        & 3838  & --    & --    & -- \\
    10187   & earthquake    & 1.90  & 1.1995        & 2733  & --    & --    & -- \\
    10188   & evil  & 1.90  & 1.2817        & 975   & 1416  & --    & 781 \\
    10189   & wars  & 1.90  & 1.3286        & 1654  & 3252  & 4696  & 2888 \\
    10190   & abuse & 1.88  & 1.2395        & 2809  & 2865  & 2236  & 3069 \\
    10191   & diseases      & 1.88  & 0.9398        & --    & 2307  & 4795  & -- \\
    10192   & sadness       & 1.88  & 1.1891        & --    & --    & 3820  & 1930 \\
    10193   & violence      & 1.86  & 1.0500        & 4299  & 1724  & 1238  & 2016 \\
    10194   & cruel & 1.84  & 1.1493        & 2963  & --    & --    & 1447 \\
    10195   & cry   & 1.84  & 1.2835        & 1028  & 3075  & --    & 226 \\
    10196   & failed        & 1.84  & 0.9971        & 2645  & 1618  & 1276  & 2920 \\
    10197   & sickness      & 1.84  & 1.1843        & 4735  & --    & --    & 3782 \\
    10198   & abused        & 1.83  & 1.3101        & --    & --    & --    & 4589 \\
    10199   & tortured      & 1.82  & 1.4241        & --    & --    & --    & 4693 \\
    10200   & fatal & 1.80  & 1.5253        & --    & 4089  & --    & 3724 \\
    10201   & killings      & 1.80  & 1.5386        & --    & --    & 4914  & -- \\
    10202   & murdered      & 1.80  & 1.6288        & --    & --    & --    & 4796 \\
    10203   & war   & 1.80  & 1.4142        & 468   & 175   & 291   & 462 \\
    10204   & kills & 1.78  & 1.2337        & 2459  & --    & --    & 2857 \\
    10205   & jail  & 1.76  & 1.0214        & 1642  & --    & 2573  & 1619 \\
    10206   & terror        & 1.76  & 1.0012        & 4625  & 4117  & 4048  & 2370 \\
    10207   & die   & 1.74  & 1.1920        & 418   & 730   & 2605  & 143 \\
    10208   & killing       & 1.70  & 1.3590        & 1507  & 4428  & 1672  & 998 \\
    10209   & arrested      & 1.64  & 1.0053        & 2435  & 4474  & 1435  & -- \\
    10210   & deaths        & 1.64  & 1.1386        & --    & --    & 2974  & -- \\
    10211   & raped & 1.64  & 1.4251        & --    & --    & --    & 4528 \\
    10212   & torture       & 1.58  & 1.0515        & 3175  & --    & --    & 3126 \\
    10213   & died  & 1.56  & 1.1980        & 1223  & 866   & 208   & 826 \\
    10214   & kill  & 1.56  & 1.0529        & 798   & 2727  & 2572  & 430 \\
    10215   & killed        & 1.56  & 1.2316        & 1137  & 1603  & 814   & 1273 \\
    10216   & cancer        & 1.54  & 1.0730        & 946   & 1884  & 796   & 3802 \\
    10217   & death & 1.54  & 1.2811        & 509   & 307   & 373   & 433 \\
    10218   & murder        & 1.48  & 1.0150        & 2762  & 3110  & 1541  & 1059 \\
    10219   & terrorism     & 1.48  & 0.9089        & --    & --    & 3192  & -- \\
    10220   & rape  & 1.44  & 0.7866        & 3133  & --    & 4115  & 2977 \\
    10221   & suicide       & 1.30  & 0.8391        & 2124  & 4707  & 3319  & 2107 \\
    10222   & terrorist     & 1.30  & 0.9091        & 3576  & --    & 3026  & -- \\
    \hline
  \end{tabular}
  \caption{
    The 50 most negative words in our data set.
  }
  \label{tab:wordhap.sadwords}
\end{table*}

\begin{table*}[htbp]
  \centering
  \begin{tabular}{rrrrrrrr}
    \hline
    $h_{\rm rank}$ & word  & $h_{\rm avg}$ & $h_{\sigma}$ & TW rank & GB rank & NYT rank  & ML rank \\
    \hline
    \hline
    8426    & fucking       & 4.64  & 2.9260        & 448   & --    & --    & 620 \\
    9263    & fuckin        & 3.86  & 2.7405        & 1077  & --    & --    & 688 \\
    9469    & fucked        & 3.56  & 2.7117        & 1840  & --    & --    & 904 \\
    8020    & pussy & 4.80  & 2.6650        & 2019  & --    & --    & 949 \\
    3770    & whiskey       & 5.72  & 2.6422        & --    & --    & --    & 2208 \\
    9462    & slut  & 3.57  & 2.6300        & --    & --    & --    & 4071 \\
    9652    & cigarettes    & 3.31  & 2.5997        & --    & --    & --    & 3279 \\
    9043    & fuck  & 4.14  & 2.5794        & 322   & --    & --    & 185 \\
    8797    & mortality     & 4.38  & 2.5546        & --    & 3960  & --    & -- \\
    9767    & cigarette     & 3.09  & 2.5163        & --    & --    & --    & 2678 \\
    10050   & motherfuckers & 2.51  & 2.4675        & --    & --    & --    & 1466 \\
    3801    & churches      & 5.70  & 2.4599        & --    & 2281  & --    & -- \\
    9985    & motherfucking & 2.64  & 2.4558        & --    & --    & --    & 2910 \\
    6390    & capitalism    & 5.16  & 2.4524        & --    & 4648  & --    & -- \\
    9015    & porn  & 4.18  & 2.4302        & 1801  & --    & --    & -- \\
    1516    & summer        & 6.40  & 2.3905        & 896   & 1226  & 721   & 590 \\
    2914    & beer  & 5.92  & 2.3891        & 839   & 4924  & 3960  & 1413 \\
    9759    & execution     & 3.10  & 2.3889        & --    & 2975  & --    & -- \\
    1830    & wines & 6.28  & 2.3737        & --    & --    & 3316  & -- \\
    9179    & zombies       & 4.00  & 2.3733        & 4708  & --    & --    & -- \\
    8898    & aids  & 4.28  & 2.3477        & 2983  & 3996  & 1197  & -- \\
    7839    & capitalist    & 4.84  & 2.3418        & --    & 4694  & --    & -- \\
    9370    & revenge       & 3.71  & 2.3363        & --    & --    & --    & 2766 \\
    2716    & mcdonalds     & 5.98  & 2.3342        & 3831  & --    & --    & -- \\
    1400    & beatles       & 6.44  & 2.3313        & 3797  & --    & --    & -- \\
    8348    & islam & 4.68  & 2.3250        & --    & 4514  & --    & -- \\
    5785    & pay   & 5.30  & 2.3234        & 627   & 769   & 460   & 499 \\
    6205    & alcohol       & 5.20  & 2.3212        & 2787  & 2617  & 3752  & 3600 \\
    9818    & muthafuckin   & 3.00  & 2.3094        & --    & --    & --    & 4107 \\
    2145    & christ        & 6.16  & 2.3067        & 2509  & 909   & 4238  & 1526 \\
    10016   & motherfuckin  & 2.58  & 2.3043        & --    & --    & --    & 1562 \\
    2074    & burger        & 6.18  & 2.3008        & 2070  & --    & --    & -- \\
    6931    & thunder       & 5.06  & 2.2983        & 3681  & --    & --    & 1313 \\
    9592    & whores        & 3.40  & 2.2946        & --    & --    & --    & 4275 \\
    3016    & naked & 5.90  & 2.2879        & 1317  & 4908  & --    & 1343 \\
    4347    & \#iphone      & 5.58  & 2.2865        & --    & --    & --    & -- \\
    5481    & liquor        & 5.36  & 2.2836        & 4915  & --    & --    & 2372 \\
    9553    & radiation     & 3.45  & 2.2827        & --    & 2847  & --    & -- \\
    8416    & wolves        & 4.65  & 2.2781        & --    & --    & --    & 3835 \\
    8511    & recall        & 4.60  & 2.2768        & 4770  & 3177  & 4105  & 1950 \\
    5625    & walmart       & 5.34  & 2.2733        & 2817  & --    & --    & -- \\
    7414    & socialism     & 4.96  & 2.2727        & --    & 4605  & --    & -- \\
    961     & marriage      & 6.70  & 2.2700        & 2444  & 1050  & 1246  & -- \\
    9882    & bombs & 2.86  & 2.2679        & --    & --    & --    & 2867 \\
    2920    & christianity  & 5.92  & 2.2663        & --    & 2554  & --    & -- \\
    4549    & vodka & 5.56  & 2.2602        & 3606  & --    & --    & -- \\
    8420    & crazy & 4.64  & 2.2566        & 383   & --    & 4761  & 312 \\
    5345    & sushi & 5.40  & 2.2497        & 2232  & --    & --    & -- \\
    3385    & god's & 5.80  & 2.2497        & --    & 1915  & --    & -- \\
    9251    & drunk & 3.88  & 2.2464        & 1006  & --    & --    & 1140 \\
    \hline
  \end{tabular}
  \caption{
    The top 50 words according to the standard deviation of happiness estimates.
  }
  \label{tab:wordhap.controversialwords}
\end{table*}

\end{document}